\documentstyle[aps,12pt,manuscript]{revtex}
\begin{document}
\preprint{INJE-TP-98-8}

\title{Greybody Factors in the $AdS_3$/CFT Correspondence}

\author{ H.W. Lee and Y. S. Myung}
\address{Department of Physics, Inje University, Kimhae 621-749, Korea} 

\maketitle

\begin{abstract}
We propose a new method to calculate the greybody factor in the 
$AdS_3$. This is based on both the non-normalizable modes of a test 
field($\Phi_i$) and $AdS_3$/CFT correspondence. 
Such non-normalizable modes serve 
as classical, non-fluctuating background and encode the choice of 
operator insertion(${\cal O}^i$) in the boundary. 
Actually specifying the boundary 
condition at infinity of $AdS_3$ corresponds to turning on these 
non-normalizable modes. Hence we can calculate the greybody factor of 
test fields without the Dirichlet(or Neumann) boundary 
condition. The result is consistent with those of the 
boundary CFT and effective string calculations. 
\end{abstract}
%\vfill
%Compiled at \today.

\newpage

Recently the $AdS$/CFT correspondence has attracted much interest
\cite{Mal9711200,Gub9802109,Wit9802150}. 
This is based on the duality relation between the string theory 
on $AdS_{d+1} \times M$ and a conformal field theory living 
on the d-dimensional boundary(${\cal B}$) of the $AdS$ sector. The relevant 
relation between string theory in the bulk and field theory 
on the boundary is 
\begin{equation}
{\cal Z}_{\rm eff}(\Phi_i) = e^{i S_{\rm eff}(\Phi_i)} =
\langle T e^{i \int_{\cal B} \Phi_{b,i} {\cal O}^i} \rangle,
\label{eq_partition}
\end{equation}
where $S_{\rm eff}$ is the effective action in the bulk, 
$\Phi_{b,i}$ is the test field $\Phi_i$ on the boundary, and $T$ is the 
time-ordered product in the boundary. 
There is a one-to-one correspondence between the local operator(${\cal O}^i$) 
of the boundary CFT and the bulk(test) field($\Phi_i$) in $AdS_3$.
For example, the test fields$\{\Phi_i\}$ are those
in the D1-D5 brane system: a free
scalar($\phi$) which couples to (1,1) operator in the boundary; two fixed
scalars ($\nu, \lambda$) to (2,2), (3,1), and (1,3) 
operators\cite{Cal97NPB65,Lee9708099}; two
intermediate scalars ($\eta, \xi$) to (1,2) and (2,1) 
operators\cite{Kle97NPB157}.
The expectation value $\langle \cdots \rangle$ 
is taken in the CFT with $\Phi_{b,i}$ as a source. From the boundary CFT, 
we can derive the bulk equation of $\Phi_i$ with mass($m$) 
and spin($s$)\cite{Mal9804085,Boe9806104}.  
Here we take $AdS_3 \times S^3$ as a relevant model. 
This is so because the (D-brane) 5D black hole becomes $AdS_3 \times S^3$ 
near horizon. Using the $AdS_3$/CFT correspondence, one can obtain 
much information of the dynamical aspects(greybody factor = 
absorption cross section) of this black hole.  Actually the 
$AdS_3 \times S^3$ is an exact solution of string theory and there is 
an exact CFT on its boundary at spatial infinity\cite{Gub97PRD7854}.

Gubser has derived the general formular for the greybody factor 
by using the effective string model\cite{Gub97PRD7854}:
\begin{eqnarray}
\sigma_{\rm abs} &=& 
{ {2 (h_R + h_L -1 )^2 (2 \pi l T_L)^{2h_L-1} 
       (2 \pi l T_R)^{2 h_R -1} } \over 
  { \omega \Gamma(2 h_L) \Gamma(2 h_R) } } \nonumber \\
&&~~\times
\sinh{\left ({\omega \over 2T_H }\right )} \left \vert 
\Gamma(h_L + i {\omega \over {2 \pi T_L}} )
\Gamma(h_R + i {\omega \over {2 \pi T_R}} ) \right \vert^2.
\label{abs_teo}
\end{eqnarray}
Also Teo recovered the same form by using the boundary CFT\cite{Teo9805014}. 
In the boundary CFT calculation,
the Poincar\'{e} coordinates 
($ds^2/l^2 = (-d\tilde t^2 +d\tilde x^2 + dz^2)/z^2 + d\Omega_3^2$)
are used for studying the region from the $z=l$ to the near horizon 
($z=\infty$).  
The boundary was taken to be at
$z=l$ rather than infinity($z=0$).  He argued that this choice 
bypasses the problem of field, 
which becomes divergent at infinity of $AdS_3$.  
The key information was encoded in the two-point function for ${\cal O}^i$,
\begin{equation}
\langle {\cal O}({\bf X}) {\cal O}({\bf Y}) \rangle = 
{ { 2 (h_L+h_R-1)^2 l^{2(h_L+h_R)-3}} \over 
  { \pi \left \vert {\bf X} - {\bf Y} \right \vert^{2(h_L+h_R)} } }.
\label{green}
\end{equation}

In this paper, we calculate the greybody factor of a set of test fields 
by using their non-normalizable mode and $AdS_3$/CFT correspondence. The 
normalizable and non-normalizable modes emerge natually either from 
a direct solution of the wave equation (\ref{eq_scalar}) or from the field 
representation of the $AdS_3$ isometry group 
($SL(2,R)_L \times  SL(2,R)_R$)\cite{Bal9805171}. The former propagates in the 
bulk and corresponds to physical state, whereas the latter palys a role of 
classical, non-fluctuating background. 
We stress that the non-normalizable mode encodes 
the choice of operator insertion in the boundary.  In the conventional 
study of $AdS_3$, the boundary condition is crucial for obtaining 
the sensible result\cite{Cha97PRD7546}.  
This is so because the $AdS_3$ is not globally hyperbolic, so that 
information may enter or exit from the boundary at infinity. 
There are the Dirichlet or Neumann boundary condition 
at infinity. If one requires the Dirichlet condition, the cross section 
can be calculated only for $m^2l^2 < -3/4$\cite{Lee9803227}.  
Further the stability on the $AdS_3$ 
requires $m^2l^2 \ge -1$\cite{Wit9802150}.  
Hence if one takes the stability and the Dirichlet condition 
seriously, the conventional method can be applied for 
the test field belonging to 
$-1 \le m^2 l^2 < -3/4$. Here includes the tachyon with 
$m^2 l^2 =-1$\cite{Mal9804085}, which couples to minimal 
weight primary operator 
(1/2,1/2).  
However, the relevant fields$\{\Phi_i\}$ are a free scalar 
field($\phi$) with $m^2l^2=0$, fixed scalars ($\nu, \lambda$) with 
$m^2l^2=8$, and intermediate scalars ($\eta, \xi$) with $m^2l^2=3$. 
Those remain unsolved in the conventional approach. 
Hence we have to develope a new calculation scheme to cover 
the relevant fields.  This is just our method based on the non-normalizable 
modes\cite{Lee9803080,Lee9804095,Lee9805050}.  In this case, specifying 
the boundary condition at spatial infinity corresponds to introducing 
the non-normalizable modes\cite{Bal9805171}. 
That is, the non-normalizable solution 
to the equation of motion corresponds to an operator insertion at 
infinity and a particular choice of boundary conditions. Hence, 
using this method, one can calculate the greybody factors of 
relevant fields without the additional Dirichelt or Neumann boundary condition.

We start with the effective action for a test field $\Phi$ 
with mass $m$\cite{Wit9802150,Bal9805171}
\begin{equation}
S_{\rm eff} = {1 \over 2}\int_{AdS_3} d^3 x \sqrt{-g} \left [
   \left ( \nabla \Phi \right )^2 + m^2 \Phi^2 \right ], 
\label{action}
\end{equation}
where $m^2=(h_L+ h_R)(h_L + h_R -2)/l^2$ according to the 
$AdS_3$/CFT correspondence\cite{Wit9802150,Boe9806104,Bal9805171}. 
The equation of motion leads to
\begin{eqnarray}
\nabla^2 \Phi  - m^2 \Phi^2   &=& 0.
\label{eq_scalar}  
\end{eqnarray}
The BTZ black hole background(locally, $AdS_3$) is given by\cite{Ban92PRL1849}
\begin{eqnarray}
 \bar g_{\mu\nu} =
 \left(  \begin{array}{ccc} - (M - {r^2 / l^2}) & -{J / 2} & 0  \\
                             -{J / 2} & r^2 & 0  \\
    0 & 0 & f^{-2}
         \end{array}
 \right)
\label{bck_metric}
\end{eqnarray}
with $f^2 =r^2 / l^2 -M +J^2 / 4 r^2$.
The metric $\bar g_{\mu\nu}$ is singular at $r=r_{\pm}$,
\begin{equation}
r_{\pm}^2 = {{Ml^2} \over 2} \left \{ 1 \pm \left [ 
   1 - \left ( {J \over Ml} \right )^2 \right ]^{1/2} \right \}
\label{horizon}
\end{equation}
with $M=(r_+^2 + r_-^2) / l^2, J=2 r_+r_- / l$.
Here $l \gg r_+ > r_-$. 
For convenience, we list the Hawking temperature $T_H$, the area of 
horizon ${\cal A}_H$, and the angular velocity at the horizon
$\Omega_H$ as
\begin{equation}
T_H = (r_+^2 - r_-^2) / 2 \pi l^2 r_+,
~~{\cal A}_H = 2 \pi r_+,
~~\Omega_H = J / 2 r_+^2.
\label{temp}
\end{equation}

Considering the $t$ and $x$-translational symmetries of the background 
metric (\ref{bck_metric}),
one chooses the perturbation for a test field as
\begin{eqnarray}
\Phi(t,\phi,r)&&=e^{-i \omega t} e^{i \mu \phi} \tilde\Phi(r).
\label{ptr_scalar}
\end{eqnarray}
Hence (\ref{eq_scalar}) becomes the equation\cite{Lee9803080}
\begin{eqnarray}
\left [ f^2 \partial_r^2 
+ \left\{ {1 \over r} (\partial_r rf^2) \right\} \partial_r
-{{J \mu \omega} \over {r^2 f^2}} 
+{\omega^2 \over f^2} 
+{{M-{r^2 \over l^2}} \over {r^2 f^2}} \mu^2 
\right ] \tilde \Phi
-m^2 \tilde \Phi 
=0,
\label{eq_decoupled}
\end{eqnarray}

Now we calculate the absorption cross section of the test field 
to obtain its dynamical information from the $AdS_3$/CFT correspondence.
Since it is hard to find a 
solution to (\ref{eq_decoupled}) directly, 
we use a matching procedure. 
The spacetime is divided into two regions: the near region 
($r \sim r_+ \ll l$) 
and far region ($r \gg l$).  We now study each region in turn.
For the far region($r \gg l$), the equation (\ref{eq_decoupled}) 
becomes
\begin{equation}
\tilde \Phi_{\rm far}'' + {3 \over r} \tilde \Phi_{\rm far}' 
- {\Lambda \over r^2} \tilde \Phi_{\rm far}=0.
\label{eq_far}
\end{equation}
Here $\Lambda=m^2 l^2 + \epsilon$ is introduced with the 
small parameter $\epsilon$ for the technical reason.
First we find the far region solution
\begin{equation}
\tilde \Phi_{\rm far}(x) = 
{1 \over x} \left ( \alpha x^{\sqrt{1+\Lambda}} 
                  + \beta x^{-\sqrt{1+\Lambda}} \right ) 
= \tilde \Phi_{\rm far}^{\rm non-nor} 
  +\tilde \Phi_{\rm far}^{\rm nor}
\label{sol_far}
\end{equation}
with two unknown coefficients $\alpha$, $\beta$ and $x=r/l$.
We note that for the case of $m^2l^2 \ge 0$, 
the first(second) term diverges(converges) at $x=\infty$, 
and thus turn out to be the non-normalizable(normalizable) 
modes\cite{Bal9805171}.  
We then calculate the ingoing flux at infinity and this is given by
\begin{equation}
{\cal F}_{\rm in}(\infty) = -2 \pi \sqrt{1+\Lambda} 
\left \vert \alpha - i \beta \right \vert^2.
%\nonumber
\label{in_flux}
\end{equation}
Here we need both non-normalizable and normalizable modes to obtain 
the non-zero incoming flux at infinity.
In order to obtain the near region behavior,
we introduce the variable 
$z={{r^2 - r_+^2} \over {r^2 - r_-^2}}=
{{x^2 - x_+^2} \over {x^2 - x_-^2}},~~0 \le z \le 1$.
Then Eq.(\ref{eq_decoupled}) leads to
\begin{equation}
z(1-z) {{d^2 \tilde \Phi} \over dz^2}
+(1-z) {{d \tilde \Phi} \over d z}
+\left ( {A_1 \over z} -{{\Lambda/4} \over 1-z} +B_1 \right ) \tilde \Phi
=0,
\label{eq_hyper}
\end{equation}
where 
$A_1 = 
\left ( {{\omega - \mu \Omega_H} \over 4 \pi T_H} \right )^2, 
B_1 = - {r_-^2 \over r_+^2}
\left ({{\omega - \mu \Omega_H r_+^2 / r_-^2} \over 
4 \pi T_H} \right )^2$.
The solution for Eq.(\ref{eq_hyper}) is given by the hypergeometric 
functions
\begin{eqnarray}
\tilde \Phi_{\rm near}(z) &=&
C_1 z^{-i \sqrt{A_1}} (1 -z )^{(1 - \sqrt{1+\Lambda})/2} F(a,b,c;z)
\nonumber \\
&&~~~~~~~~+C_2 z^{i \sqrt{A_1}} (1 -z )^{(1 - \sqrt{1+\Lambda})/2} 
          F(b-c+1,a-c+1,2-c;z),
\label{sol_hyper}
\end{eqnarray}
where
\begin{eqnarray}
a&=& \sqrt{B_1} - i \sqrt{A_1} +(1 - \sqrt{1+\Lambda} )/2,
\nonumber \\
b&=& - \sqrt{B_1} - i \sqrt{A_1} + (1 - \sqrt{1+\Lambda} )/2,
\nonumber \\
c&=& 1 - 2 i \sqrt{A_1}.
\nonumber 
\end{eqnarray}
and $C_1$ and $C_2$ are to-be-determined constants.
At the near horizon($r\sim r_+, z \sim 0$), (\ref{sol_hyper})
becomes
\begin{eqnarray}
\tilde \Phi_{\rm near}(0) &\simeq& C_1 z^{-i \sqrt{A_1}} + C_2 z^{i \sqrt{A_1}}
\nonumber \\
&=&C_1 \left ( { 2 x_+ \over {x_+^2- x_-^2}} \right ) ^{-i \sqrt{A_1}}
           e^{-i \sqrt{A_1} \ln(x-x_+)}
+C_2 \left ( { 2 x_+ \over {x_+^2- x_-^2}} \right ) ^{i \sqrt{A_1}}
           e^{i \sqrt{A_1} \ln(x-x_+)}.
\label{sol_z0}
\end{eqnarray}
Taking an ingoing mode at horizon, we have $C_2=0$. 
Hence the near region solution is
\begin{eqnarray}
\tilde \Phi_{\rm near}(z) &=&
C_1 z^{-i \sqrt{A_1}} (1 -z )^{(1 - \sqrt{1+\Lambda})/2} F(a,b,c;z).
\label{sol_near}
\end{eqnarray}
Now we need to match the far region solution (\ref{sol_far}) to the large 
$r(z\to 1$) limit  of near region solution (\ref{sol_near}) in the 
overlapping region.  The $z\to 1$ behavior of (\ref{sol_near}) follows 
from the $z \to 1-z$ transformation rule for hypergeometric functions. 
Using $1-z \sim (x_+^2 - x_-^2)/x^2$ for $r \to \infty$, 
this takes the form
\begin{eqnarray}
\tilde \Phi_{n\to f}(r) &\simeq&
C_1 E_1
{x^{ \sqrt{1+\Lambda}}\over x} 
+C_1 E_2 
{x^{-\sqrt{1+\Lambda}}\over x},
\label{sol_z}
\end{eqnarray}
where
\begin{eqnarray}
E_1 &=& 
{{ \Gamma(1 -2 i \sqrt{A_1}) \Gamma(\sqrt{1+\Lambda})
    (x_+^2- x_-^2)^{(1-\sqrt{1+\Lambda})/2}} \over
 {\Gamma({{1+\sqrt{1+\Lambda}} \over 2} + \sqrt{B_1} -i \sqrt{A_1} ))
  \Gamma({{1+\sqrt{1+\Lambda}} \over 2} - \sqrt{B_1} -i \sqrt{A_1} ))}},
\label{ain} \\
E_2 &=&
{{ \Gamma(1 -2 i \sqrt{A_1}) \Gamma(- \sqrt{1+\Lambda})
   (x_+^2-x_-^2)^{(1+ \sqrt{1+\Lambda})/2}} \over
  {\Gamma({{1-\sqrt{1+\Lambda}} \over 2} + \sqrt{B_1} -i \sqrt{A_1} ))
   \Gamma({{1-\sqrt{1+\Lambda}} \over 2} - \sqrt{B_1} -i \sqrt{A_1} ))}}.
\label{aout}
\end{eqnarray}
Matching (\ref{sol_far}) with  (\ref{sol_z}) 
leads to $\alpha = C_1 E_1$ and $\beta = C_1 E_2$.  Considering 
$x_+^2-x_-^2 \ll 1 $ and $m^2l^2 \ge 0$, then one finds $\beta \ll \alpha$.  
Thus the ingoing flux across the horizon is given by
${\cal F}_{\rm in}(0) = -8 \pi \sqrt{A_1} (x_+^2- x_-^2) \vert C_1 \vert^2$.
This amounts to taking the flux of the non-normalizable mode effectively. 
At this stage, we wish to comment on the parameter $\Lambda$.  
Initially we introduce
$\Lambda=m^2l^2+\epsilon$ with the small parameter $\epsilon$ in
Eq.(\ref{eq_far}).  This is so because $E_2$ in (\ref{aout})
has a pole for integral $\Lambda$ through $\Gamma(-\sqrt{1+\Lambda})$.  
Hence it is convenient to keep $\Lambda$ near an integer 
value during the calculation and make it integer at the end.
Hence for $\mu=0, \epsilon=0(\Lambda=m^2 l^2)$, 
we can obtain the absorption coefficient 
\begin{eqnarray}
&&{\cal A} = 
{{\cal F}_{\rm in}(0) \over {\cal F}_{\rm in}(\infty)} = 
{ 4 \sqrt{A_1} (x_+^2-x_-^2) \over \sqrt{1+m^2 l^2} }
{1 \over |E_1|^2}
\nonumber \\
&&=
{\omega {\cal A}_H \over \pi} 
{(x_+^2-x_-^2)^{(\sqrt{1+m^2 l^2}-1)} \over 
{\Gamma(1+\sqrt{1+m^2 l^2})\Gamma(\sqrt{1+m^2 l^2})}}
\left \vert 
{{\Gamma({{1+\sqrt{1+m^2 l^2}} \over 2} - i { \omega \over 4 \pi T_L})
  \Gamma({{1+\sqrt{1+m^2 l^2}} \over 2} - i { \omega \over 4 \pi T_R})}
\over
  \Gamma(1 -i {\omega \over 2 \pi T_H})}
\right \vert^2,
\label{cross} 
\end{eqnarray} 
where left and right temperatures are defined by
\begin{equation}
{1 \over T_{L/R}} = {1 \over T_H} \left ( 1 \pm {r_- \over r_+} \right ).
\label{temperature}
\end{equation}
The absorption corss-section is given by 
$\sigma_{\rm abs} = {\cal A} / \omega$ in three dimensions.
For $h_L=h_R$, this takes the form 
\begin{eqnarray}
{ \sigma_{\rm abs}^{AdS} }
&=& {{2 (h_L + h_R -1) (2 \pi l T_L)^{2h_L-1}(2 \pi l T_R)^{2h_R-1} } \over 
{\pi \omega \Gamma(2h_L)\Gamma(2h_R)}} 
\sinh{\left({\omega \over 2 T_H}\right )}  
\nonumber \\
&&~~~~\times
\left \vert 
{\Gamma(h_L - i { \omega \over 4 \pi T_L})
  \Gamma(h_R - i { \omega \over 4 \pi T_R})}
\right \vert^2.
\label{abs_ads}
\end{eqnarray}
This is our key result.
In the $m^2l^2 \to 0$ limit, (\ref{abs_ads}) recovers the decay rate for 
the free scalar$(\phi)$ which couples to (1,1) operator
\cite{Lee9804095,Bir97PLB281} 
\begin{equation}
\Gamma_{\rm min} 
= { \sigma_{\rm abs}^{\rm min} \over {e^{\omega \over T_H} -1 }}
= {{\pi l^2 \omega} \over 
 {(e^{\omega \over 2 T_R} -1) 
  (e^{\omega \over 2 T_L} -1)}}.
\label{decay_min}
\end{equation}
It is pointed out that the dilaton as a fixed scalar$(\nu)$ which 
couples to (2,2) is physically propagating field in the BTZ back ground.  
In the limit of $m^2l^2 \to 8$, our result (\ref{abs_ads}) is 
that for the dilaton as was shown in Refs.\cite{Lee9803080,Lee9805050}. 
For $m^2l^2=3$, one can obtain the cross section for a new  
scalar which couples 
to the primary operator (3/2,3/2)\cite{Mal9804085}.

In this $AdS_3$-calculation, there exists an 
ambiguity in determining normalization 
factor of the cross section for test field with $m^2l^2 > 0$ 
in comparison with the previous methods.  In the case 
of semiclassical calcualtion with asymptotically flat space, 
there is no ambiguity in calculating the 
cross section\cite{Cal97NPB65,Lee9708099}. 
This includes both $M_5 \times S^1 \times T^4$ (5D black hole) and 
$AdS_3 \times S^3 \times T^4$ near horizon with the asymptotically 
flat space\cite{Lee9804095,Lee9805050}.  
Also, in the effective string approach one can calculate 
the cross section without arbitrariness\cite{Cal97NPB65,Gub97PRD7854}.  
However, as in the effective CFT 
method by Maldacena and Strominger\cite{Mal97PRD4975}, 
it seems that there is no known 
way to fix the normalization factor without appealing to string theory. 
To recover the effective string and boundary CFT 
results (\ref{abs_teo}) correctly, 
we need a conversion factor ($\pi(h_L+h_R-1)$). 
As a definite example, we list the cross section of the fixed scalar$(\nu)$ 
which couples to (2,2) operator. This is 
$\sigma_{\rm abs}^{\omega\to 0} = C {\cal A}_H ( r_0/l)^4$. 
Here $C=1/4$ for the semiclassical calculation, effective string method, 
boundary CFT-calculation in (\ref{abs_teo}), whereas $C=1/12$ for our 
$AdS_3$-calculation. However, there is no ambiguity for a free 
scalar with $m^2l^2=0$.

Now we are in a position to discuss other fields which couple to 
chiral operators as (3,1), (1,3), (2,1), (1,2), (2,0), (0,2), (1,0), and 
(0,1).  We remind the reader that the fixed scalar($\lambda$) couples to 
(3,1) and (1,3) as well as (2,2). This has a discrepancy in the cross section 
between 
the semiclassical and effective string 
calculations\cite{Cal97NPB65,Lee9708099}. 
Two intermediate scalars ($\eta,\xi$) couple to (1,2) 
and (2,1)\cite{Kle97NPB157}. To obtain the general form of their 
greybody factors by using the semiclassical approach 
remains as open question still.  
Also gauge bosons couple to (2,0) and (0,2) and
tachyons to (1,0) and (0,1)\cite{Mal9804085}.  Those all 
carry non-zero spin($s=h_L-h_R$) with the mass($m^2l^2=(h_L+h_R)(h_L+h_R-2)$). 
In order to study the wave equation including 
the non-zero spin, we may extract its information from 
the boundary CFT. 
We write down its equation on the boundary at the infinity as\cite{Boe9806104}
\begin{equation}
\left [ \nabla^2  + {s^2 \over {l^2 \sinh^2\rho}} - m^2 \right ]\Phi_s =0
\label{eq_global}
\end{equation}
in the global coordinates($\tau, \phi, \rho)$:
\begin{equation}
{ds^2_{\rm g} \over l^2} = 
- \cosh^2 \rho d\tau^2 + \sinh^2 \rho d\phi^2 + d\rho^2.
\label{ds_global}
\end{equation}
Its solution wiht $\nu^2 =1 + m^2l^2$ is given by
\begin{equation}
\Phi_s = e^{-i(h_L+h_R) \tau -is \phi}
\left [ {\alpha \over (\cosh \rho)^{(1-\nu)/2}} 
 + {\beta \over (\cosh \rho)^{(1+\nu)/2}} \right ]. 
\label{sol_global}
\end{equation}
We note here that the first (the second) terms correspond to the 
non-normalizable (normalizable) modes.
For large $\rho(r \to \infty)$, $\Phi_s$ vanishes. 
Near horizon it takes a plane wave. 
Introducing the Poincar\'{e} coordinates(
$\omega^+ = \tilde t + \tilde x, \omega^- = \tilde t - \tilde x, z$) 
to rewrite (\ref{eq_global}) leads to
\begin{equation}
\left [ z^2 \left ( 4 \partial_+ \partial_- + \partial^2_z 
    - {1 \over z} \partial_z \right ) 
  -{ 4 s^2 z^2 \over {z^4 -2 (1-\omega^+ \omega^-) z^2 
      + (1 + {\omega^+}^2)(1+{\omega^-}^2) } } 
  - m^2 l^2 \right ] \Phi_s = 0.
\label{eq_poincare}
\end{equation}
If $s=0$ and 
$\Phi_{s=0} = e^{-i \omega \tilde t + i\mu\tilde x}z \tilde \Phi(\eta)$ 
with $\eta = \sqrt{\omega^2 - p^2}z$, 
the above reduces to the Bessel equation
\begin{equation}
\eta^2 \ddot {\tilde \Phi} + \eta \dot{\tilde \Phi} + 
     \left [ \eta^2 - \nu^2 \right ] \tilde \Phi = 0. 
\label{eq_z_poincare}
\end{equation}
The solution for an integer $\nu=h_L+h_R-1$ is given by\cite{Mal9804085}
\begin{equation}
\Phi = e^{-i \omega \tilde t + i \mu \tilde x} 
\left [ \beta z J_\nu(\sqrt{\omega^2 - \mu^2}z)
        +\alpha z Y_\nu(\sqrt{\omega^2 - \mu^2}z) 
\right ].
\label{sol_poincare}
\end{equation}
The first term satisfies the Dirichlet boundary 
condition at $z=0(r \to \infty$) 
and corresponds to the normalizable mode.  This may be used for the 
perturbative calculation. Although the latter does not satisfy 
the Dirichlet boundary condition, it corresponds the 
non-normalizable mode to specify another boundary condition. 
And thus this mode can be used to derive the greybody factor. However, 
turning on the spin-dependent term, we have a difficulty in solving 
Eq.(\ref{eq_poincare}). 
This is mainly due to being unable to seperate those variables. 

Finally, we wish to rewrite Eq.(\ref{eq_global}) in terms of the BTZ 
coordinates($t,\phi, r$). For this purpose, we introduce the proper radial 
coordinates($\tau,\phi, \rho$)\cite{Car9806026,Beh9806195}:
\begin{equation}
ds^2_P = -\sinh^2 \rho \left ( - r_+ d\tau + r_- d \phi \right ) ^2 
   + l^2 d \rho^2 + \cosh^2 \rho \left ( 
  -r_- d\tau + r_+ d \phi \right )^2 
\label{ds_BTZ}
\end{equation}
with $r^2 = r_+^2 \cosh^2 \rho - r_-^2 \sinh^2 \rho$ and 
$\tau = t/l$ for the exterior of the BTZ black hole.  Here we define 
$\gamma$ and $\delta$ as $t$ and $\phi$,
\begin{equation}
\gamma = { {-r_- \tau + r_+ \phi  } \over l },
\delta = { {-r_+ \tau + r_- \phi } \over l }.
\label{redefine}
\end{equation}
Then Eq.(\ref{eq_global}) can be rewritten as 
\begin{equation}
\left [ l^2 \nabla^2_{\rm BTZ} 
-{ 2 s^2 (r_+^2 -r_-^2) \over 
   {(r^2 -r_-^2) \cosh{2 \gamma} +(r^2 -r_+^2) \cosh{2 \delta}
      + (r_+^2 - r_-^2) } } -m^2 l^2 \right ] \Phi_s =0.
\label{eq_BTZ}
\end{equation}
Note that the spin term depends on ($\phi, \tau$) and takes the form 
of $r^{-4}$. This is very similar 
to Eq.(\ref{eq_poincare}) which tell us that its spin term depends on
($\omega^+, \omega^-$) and takes the form of $z^{-4}$.  
Near horizon($r \to r_+$), Eq.(\ref{eq_BTZ}) takes the following form with
$x, \tau$, and $j=J/l$:
\begin{eqnarray}
&&\left [ f^2 \partial^2_x + \left \{ {1 \over x} \partial_x
    \left ( x f^2 \right ) \right \} \partial 
    - { j \over {x^2 f^2}} \partial_\phi \partial_\tau \right . \nonumber \\
&&~~~~~~~~~~
\left . - {1 \over f^2} \partial^2_\tau 
- {1 \over {f^2 x^2}} \left ( M -x^2 \right ) \partial^2_\phi
- { s^2 \over {\cosh^2(x_+\phi -x_-\tau) }} -m^2 l^2 \right ] \Phi_s =0, 
\label{eq_x_BTZ}
\end{eqnarray}
which is the same form as Eq.(\ref{eq_decoupled}) when $s=0$.  On the other 
hand, in the far region, one has
\begin{equation}
\left [ {d^2 \over dx^2} + {3 \over x} {d \over dx} 
-{{2 s^2 (x_+^2 -x_-^2) } \over 
  {x^4 (\cosh 2 \delta + \cosh 2 \gamma ) } }  
- { {m^2 l^2 } \over x^2 } \right ] \Phi_s^{\rm far} =0.
\label{eq_x_BTZ_far}
\end{equation}
When $s=0$ or $x\to \infty$, this leads to Eq.(\ref{eq_far}).
Hence, in the limit of $x\to \infty(z \to \infty)$ the spin 
term does not play an important role. Also this is confirmed from the 
observation of Eq.(\ref{eq_global}) with $\rho\to \infty$.  

In conclusion, we obtain the greybody factor for test fields$\{\Phi_i\}$ on 
the $AdS_3$.  This is based on both the non-normalizable mode of the test 
fields and $AdS_3$/CFT correspondence. Because the 
non-normalizable mode means the other type of boundary condition 
at infinity, we don't need to require the Dirichlet or Neumann condition. 
However, we wish to point out the ambiguity in determining normalization 
constant of the greybody factor in our $AdS_3$-calculation.  It seems 
that this arises from the flux (\ref{in_flux}) which is calculated in the 
asymptotically $AdS_3$. The definition of flux is clear in the 
asymptotically flat space, but its definition seems to be unclear 
in the asymptotically $AdS_3$. For the test fields which couple to 
chiral operator, there exist some discrepancies in the greybody 
factor between the semiclassical and effective string calculations. 
We may resolve this by using the $AdS_3$/CFT correspondence. 
In the above we sketch the spin-dependent wave functions briefly, but we do not 
have a significant result until now. 
On the other hand, the greybody factor in (\ref{abs_teo}) is valid 
for any $h_L$ and $h_R$. Although (\ref{abs_ads}) 
is derived for $h_L=h_R$, it takes the same form as (\ref{abs_teo}). 
Hence we believe that our result (\ref{abs_ads}) 
can be extended to $h_L \ne h_R$ case to accommodate the test field 
with non-zero spin.

\section*{Acknowledgement}
We would like to thank J.Y. Kim for usefull discussions.
This work was supported in part by the Basic Science Research Institute 
Program, Minstry of Education, Project NOs. BSRI-98-2441 and 
BSRI-98-2413.

\end{document}